\documentclass[10pt,conference]{IEEEtran}
\IEEEoverridecommandlockouts
\usepackage{preamble}
\usepackage{cite}
\usepackage{amsmath,amssymb,amsfonts}
\usepackage{algorithmic}
\usepackage{graphicx}
\usepackage{textcomp}
\usepackage{xcolor}
\def\BibTeX{{\rm B\kern-.05em{\sc i\kern-.025em b}\kern-.08em
    T\kern-.1667em\lower.7ex\hbox{E}\kern-.125emX}}
\usepackage{lipsum}

\begin{document}

\title{COFO: COdeFOrces dataset for Program Classification, Recognition and Tagging}

\author{
\IEEEauthorblockN{
Kuldeep Gautam \hspace{5mm}  
S. VenkataKeerthy \hspace{5mm} 
Ramakrishna Upadrasta
}
\{cs20mtech01004, cs17m20p100001\}@iith.ac.in, 
ramakrishna@cse.iith.ac.in
\IEEEauthorblockA{
\textit{Department of CSE,} \textit{Indian Institute of Technology Hyderabad}\\
}
}

\maketitle
\let\thefootnote\relax\footnotetext{This version was prepared in January 2021.}

\begin{abstract}
    In recent years, a lot of technological advances in computer science have aided software programmers to create innovative and real-time user-friendly software. With the creation of the software and the urging interest of people to learn to write software, there is a large collection of source codes that can be found on the web, also known as Big Code, which can be used as a source of data for driving the machine learning applications tending to solve certain software engineering problems. 

In this paper, we present COFO, a dataset consisting of 809 classes/problems with a total of 369K source codes written in C, C++, Java and Python programming languages along with other metadata such as code tags, problem specification, and input-output specifications. 
COFO has been scraped from the openly available Codeforces website using a selenium-beautifulsoup-python based scraper. 
We envision that this dataset can be useful for solving machine learning-based problems like program classification/recognition, tagging, predicting program properties, and code comprehension.

\end{abstract}


\section{Introduction}
\label{sec:introduction}
At present, lot of advancements in computer science is driven by machine learning. New ways of processing, analyzing, and understanding programs have been introduced that rely on machine learning based methods. Several applications of machine learning in the domain of software engineering and program optimizations have been proposed and widely successful~\cite{alon2019code2vec, ncc, VenkataKeerthy-2020-IR2Vec, cummins2017end2end}. These include program comprehension applications like algorithm recognition, program classification~\cite{tbcnn-aaai16, ncc}, bug detection~\cite{wang2016bugram}, 
method name prediction~\cite{alon2019code2vec, Allamanis:suggesting-acc-meth-names}, 
and summarization~\cite{bolin-cscg-neurips19, chen-bvae-ase18, alon2018codeseq, wan-RL-Summarization-ASE18}.

It is well known that the availability of a good dataset is crucial to effectively solving machine learning-based problems. Nowadays, a large amount of source code is widely available through various means and modes; this includes open-source hosting sites like GitHub, Community Question and Answer sites like Stack Overflow, and Open Judge systems. Collectively, this is often referred to as \textit{Big Code}~\cite{prog-big-code}. 
However, the original code repositories are often in a raw-form to be used effectively. They require pre-processing, curation, and labeling for ML applications.

Several datasets have been proposed and are actively being used for some of the above-mentioned applications. Allamanis et al.~\cite{conv-attn-net-extreme-summ}, and Alon et al.~\cite{alon2018codeseq} present Java based datasets crawled from GitHub for predicting meaningful method names. BigCloneBench~\cite{bigclonebench-ICSME14} is another benchmark consisting of only Java codes to identify code clones. 
Iyer et al.~\cite{iyer2016summarizing} present a dataset on C\# and SQL collected from Stack Overflow and processed for code captioning. 

POJ-104 is a well-known dataset proposed by Mou et al.~\cite{tbcnn-aaai16} for code classification tasks. It consists of 104 programming questions collected from an online judge system, which is written by students in C/C++. Each question forms a class, and each class consists of around 500 programs, forming a dataset consisting of approximately 52K programs.  To our knowledge, POJ-104 is one of the few published datasets that is available for program classification tasks. 

Another similar dataset is available from CodeJam, Google`s annual coding contest.
Several works on clone detection have crawled data from CodeJam for benchmarking. 
Works like \cite{deepsim-fse18, codeclones-saner20} collect a set of about 1,669 Java programs across 12 classes. 
Ye et al.~\cite{misim-arxiv20} form a dataset of about 297 classes of C/C++ programs from CodeJam.
It can be noticed that in comparison to the standard image classification datasets like ImageNet\footnote{\url{http://www.image-net.org/}} that has about 20K categories, these benchmarks are quite small in terms of the number of classes and programs/class and are restricted to the only a subset of languages.

There is an utmost need for a \textit{well-formed} program classification dataset that has two important properties: language independence and largeness (in number as well as variety). Meaning the dataset should have programs from different languages, with a large number of classes and a large number of programs in each class. As websites that host competitive programming have multiple problems and multiple submissions corresponding to each problem, they naturally tend to be the source of such data.

Codeforces\footnote{\url{https://codeforces.com/}} is one such online platform that hosts competitive programming contests. To our knowledge, it is one of the few sites that make the data (problems and submissions, and other information) publicly available (apparently with none of standard licenses).
It is maintained by a group of competitive programmers from ITMO University led by Mikhail Mirzayanov. This platform consists of a large amount of data in terms of source codes from various programming languages that can be extracted/scraped and used to perform many tasks such as code classification, code tagging, etc.

Existence of such platforms and the need for a good program classification dataset provide the motivation for creating a dataset of source codes across various languages, collected from a platform hosting online programming events, and to make it accessible to everyone to make use of it for driving various machine learning applications.

In this paper, we propose \textit{COFO}, a dataset consisting of \textit{369K} C/C++/Java/Python programs crawled from the codeforces platform. Inspired by the need and potential of machine learning to solve certain problems, we curated a dataset consisting of \textit{809} classes/problems as directories, with each directory consisting of a problem specification, input-output specifications, test cases, code tags and a subdirectory that contains all the C/C++/Java/Python source code submissions for that problem. 

Our process starts with the collection of metadata for each problem using a Codeforces API, followed by the generation of URLs for two different web pages for each problem. One is for scraping the specifications from the problem specification page. The second one is for crawling the source codes and test cases from a number of submission pages for the problem. The test cases for a problem have been crawled from a pop-up window that appears after clicking on the very first submission present on the submissions page. All the source codes, being accepted solutions, are compilable programs and are able to pass all the test cases defined for the problem. The dataset consists of a total of four programming languages, including a version of C (GNU C11), three versions of C++ (GNU C++11, GNU C++14, GNU C++17), two versions of Java (Java-8, Java-11), and a version of Python (Python-3).

In summary, we believe that COFO can be useful in answering the following \textbf{R}esearch \textbf{Q}uestions.

    \textbf{[Code Classification/Clone Detection] Given a collection of programs, which of them (within and across languages) solves the same problem?}
    The dataset contains several implementations of the same problem that can help in performing classification/recognition/clone detection.
    
    \textbf{[Code Tagging] Given an input program, which tags can describe the input code accurately?}
    In addition to the source codes, the dataset also contain code tags describing each problem statement, making it suitable for code tagging.
    
    \textbf{[Predicting properties] Given a program, what would be the upper bounds of memory and time for it?}
    We have also collected input-output specifications containing the constraints on inputs and outputs, such as upper bounds on size, memory, and time limit for the code. Hence we believe that the dataset can be suitable for predicting bounds on memory and time.
    
    \textbf{[Code Comprehension] What is the program trying to solve?} In addition to the discussed metadata, the dataset also contains problem specifications (textual information) describing the problem, which can be used in an NLP-based application using the source codes and the specifications together.
    

We open-source the toolchain used to collect this data at
\url{https://github.com/kgautam01/CodeForces-Scraper/}.

\section{Data Collection Methodology}

Each problem on the codeforces platform is associated with the solutions corresponding to a problem along with the following metadata.

\begin{enumerate}
    \item Problem specification: defines the problem statement.
    \item Input and output specifications: defines the constraints such as size, type, memory, and time limits of the expected input and output.
    \item Test cases: associated with each problem (one or more).
    \item Code-Tags: a collection of (mostly algorithmic) specific keywords, such as greedy, dynamic programming, graph, etc. Each problem is tagged with one or more such keywords for identification.
\end{enumerate}

Codeforces provides an API\footnote{\url{http://codeforces.com/api/problemset.problems}} for gathering the metadata about all the problems/questions, present on the website. Once a request is made through the API on the codeforces server, it returns a JSON response consisting of \textit{contestID}, \textit{Index} and \textit{Code-Tags}, where \textit{contestID} and \textit{Index} refer to the ID associated with the coding contest and the problem within the contest respectively.

Using the \textit{contestId} and \textit{index}, we construct a URL 
that serves as the base URL for crawling all the metadata or information as stated above. For each problem statement, we collect the problem specification, Input-output specifications, defined test cases, C/C++/Java/Python solutions, and respective code-tags. We make sure to crawl only the accepted solutions to the problem from the submissions. We call a solution accepted if it can compile successfully and pass all the test cases defined for that problem.

\subsection{Scraping process}

\begin{figure*}[!t]
    \centering
    \includegraphics[scale=0.12]{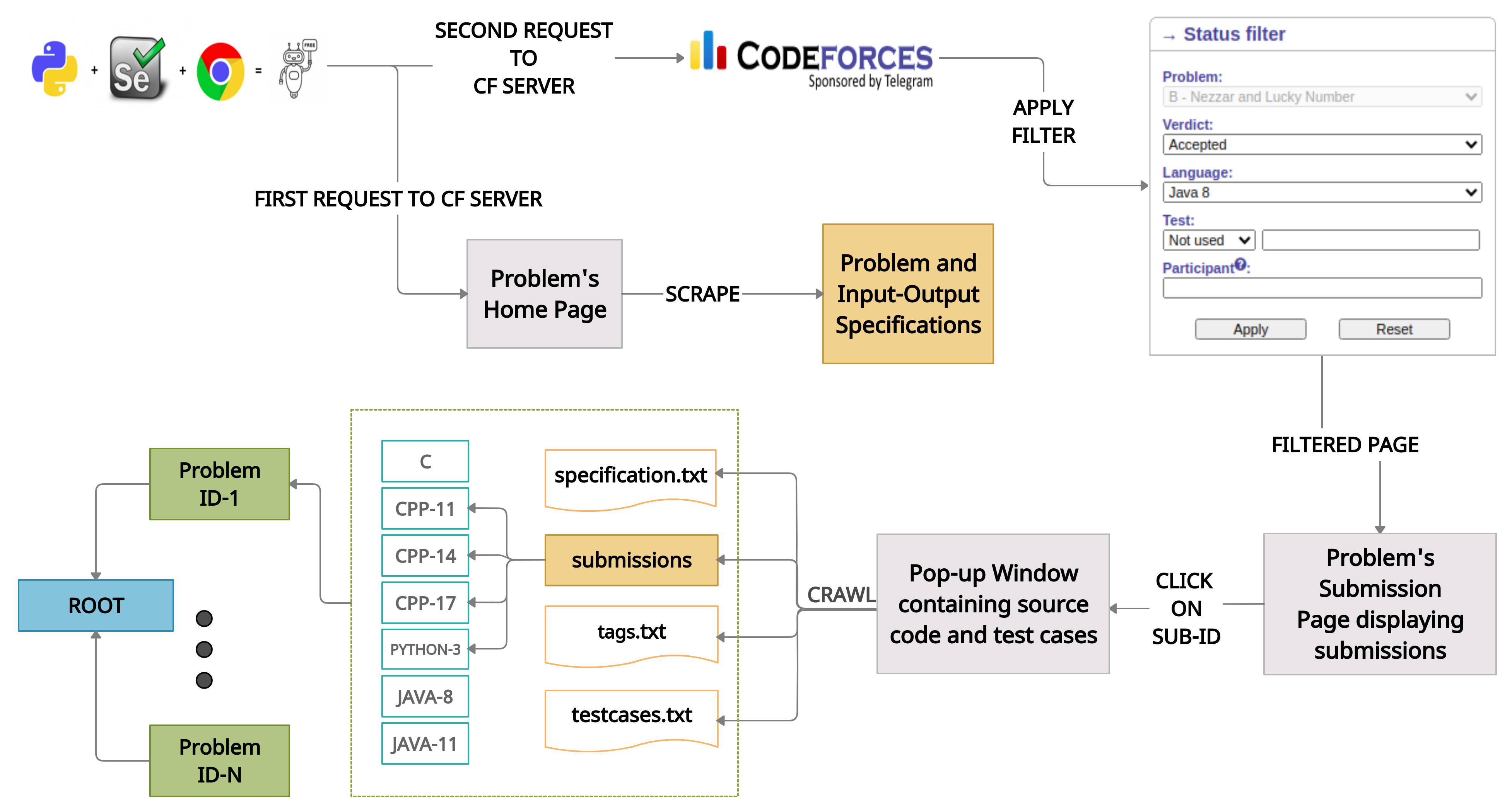}
    \caption{Schematic of the Scraping Process and the Directory Structure}
    \label{fig:scrapping-process}
\end{figure*}

The following steps summarize the data collection process for each problem statement.

\begin{enumerate}
    \item A HTTP request to the codeforces server is made with the URL associated with a problem statement, created using \textit{contestid} and \textit{index} (problem’s homepage) to scrape the problem specification, and input-output specifications using \texttt{beautifulsoup}. Once scraped, a \textit{specifications.txt} file is created in each directory. To scrape the specifications, only one request is made to the codeforces server for each problem.
    
    \item Another request to the (first) submission page for the problem is made, and filter for the language (C/C++/Python3/Java8/Java11) and accepted solutions is applied. 
    
    \item For each of the the displayed submissions on each page, the corresponding submission ID is clicked (automated using selenium) followed by the crawling of source codes from the pop-up that appears after the click event. Test cases are defined for a problem. Only the very first accepted submission is crawled to collect the test cases, keeping other iterations independent of crawling the test cases.
    
    \item The process stops once it reaches the page limit or the total number of source codes crawled per problem reaches a threshold. We fixed 750 as the threshold. Python’s logging module is used to log all the errors, if any, that occur during the process.
\end{enumerate}

Instead of issuing URL requests, we follow a selenium based approach for scraping as it reduces the number of HTTP requests. 
For instance, for a problem that has 50 submissions, scrapping them using URL requests would need 50 HTTP requests. However, with the selenium based approach it can be done with a single request utilizing that the listing on display contains a maximum of 50 submissions per page.
\section{Data Description}

\subsection{Directory Structure}

The collected data is organized in the following hierarchy.

\paragraph{Root directory} The root directory contains subdirectories, where each subdirectory corresponds to a problem. Each subdirectory is labelled with the \textit{problemID}. A \textit{problemID} is created by concatenating a \textit{contestID} with its \textit{index}.

\paragraph{Problem-dir/Subdirectory} Each subdirectory contains a submissions folder containing the acceptable solutions corresponding to the problem. The submissions folder contains a separate directory for each of the programming languages in consideration. Even for different versions of a programming language, separate directories are maintained. This is also shown in Fig.~\ref{fig:scrapping-process}.
In addition to the submissions folder, we create three text files: one (\textit{testcases.txt}) containing test cases, 
the second \textit{specifications.txt} containing the problem specification and input-output specifications, and the last one, \textit{tags.txt}, containing the tags associated with the problem.

\paragraph{Submissions} This is the innermost directory in the tree for each problem and contains the source code submissions. The corresponding submission ID is used to name these source files. As this ID is unique, it ensures that no two files have the same name, or files are not overwritten at the time of crawling or writing.

\begin{table}[!t]
\renewcommand{\arraystretch}{1.3}
\centering
\caption{Statistics of COFO Dataset}
\label{tab:statistics}
\hspace{-0.5cm}
\subfloat[\#Programs/Language]{{ 
    \begin{tabular}{|l|r|}
        \hline
        \textbf{Language} & \textbf{\#Programs} \\
        \hline
        C 11                 & 26,449     \\
        C++ 11               & 92,015     \\
        C++ 14               & 76,873     \\
        C++ 17               & 97,926     \\
        Java 8               & 33,919     \\
        Java 11              & 14,876     \\
        Python 3             & 27,044     \\
        \hline
        Total     & 369,102    \\
        \hline
    \end{tabular}
    \label{tab:lang-distribution}
}}%
\subfloat[Other statistics]{{ 
    \begin{tabular}{|l|r|}
        \hline
        \#Classes            & 809        \\
        Min \#Programs per class   & 10        \\
        Max \#Programs per class   & 2632        \\
        Avg \#Programs per class   & 456        \\
        Total \#Code-Tags   & 1955        \\
        \#Unique Code-Tags   & 35        \\
        (Min, Max) \#Code-Tags per class  & (0, 8)        \\
        \#LOC in the dataset  & 17.6M \\
        Avg \#LOC per class & 21.8K \\
        \hline
    \end{tabular}
}}%
\end{table}

\begin{figure*}[h!t]
    \centering
    \hspace{-1cm}
      \subfloat[Distribution of data points in the dataset]{{\includegraphics[scale=0.33]{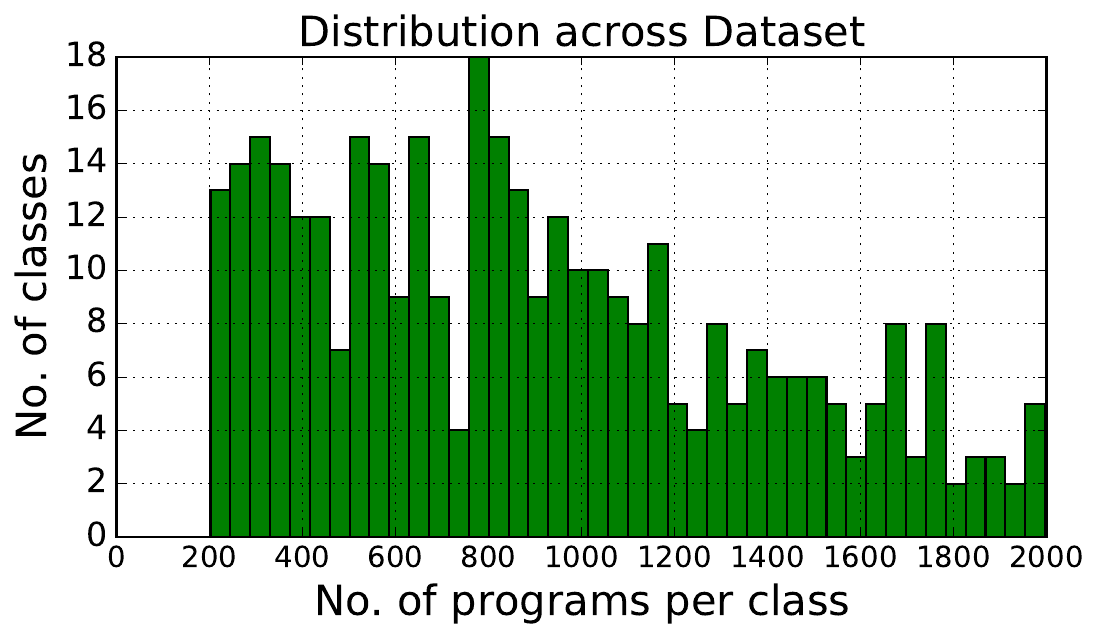} }
      \label{fig:dataset-distribution}}%
    \subfloat[{Distribution of data points per language}]{{ \includegraphics[scale=0.32]{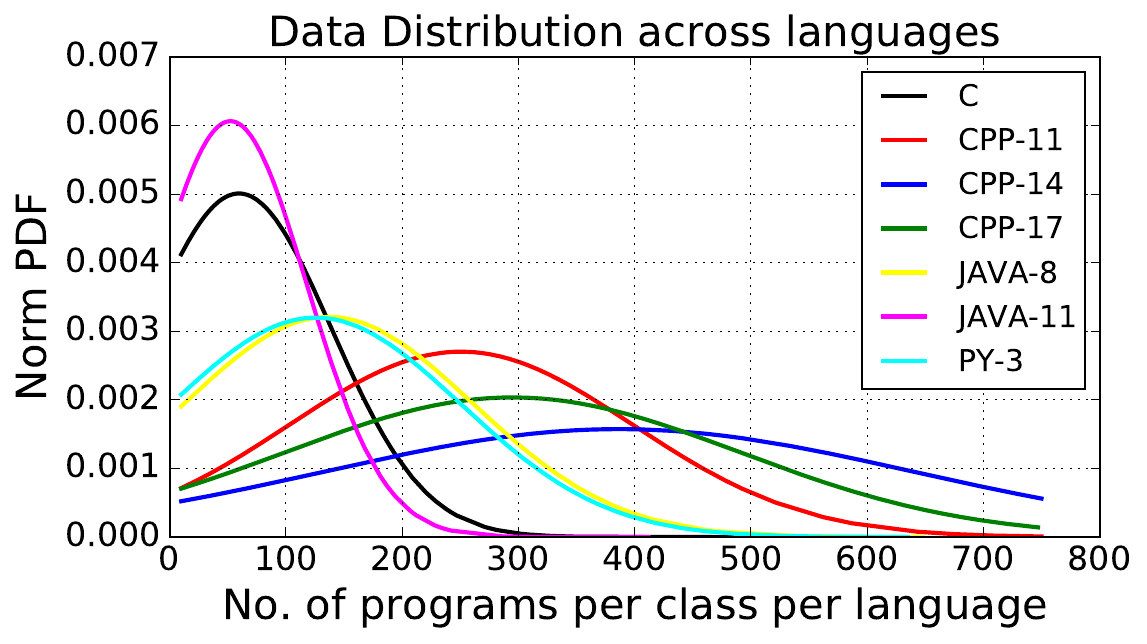}
    \label{fig:lang-distribution}}}%
    \subfloat[{Distribution of code tags in the dataset }]{{ \includegraphics[scale=0.32]{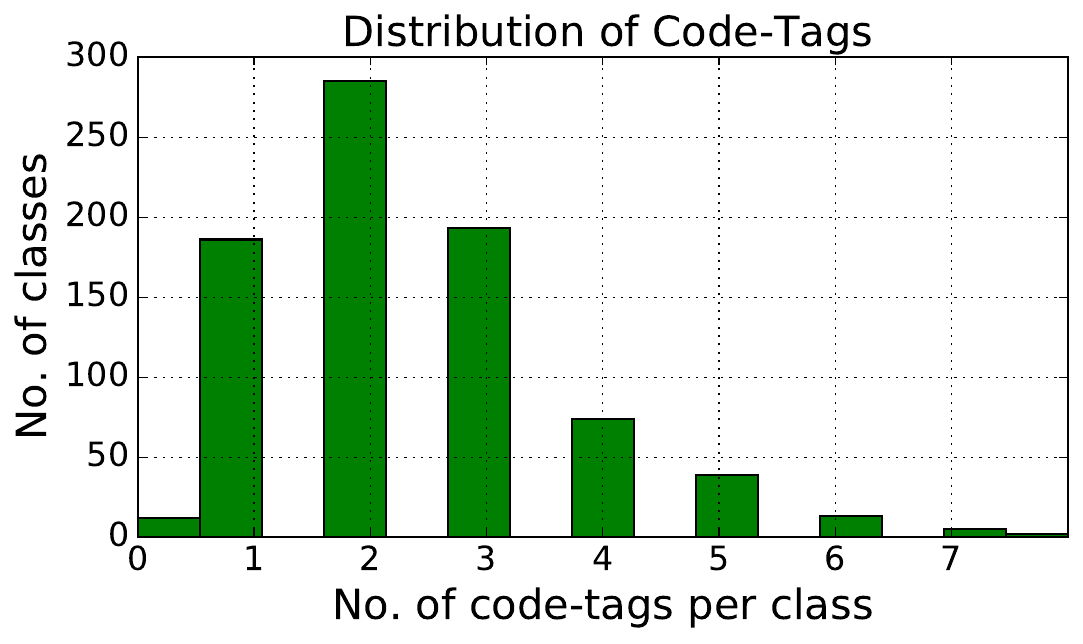} 
    \label{fig:tag-distribution}}}%
    \caption{Dataset Distribution}
    \label{fig:data-distribution}
\end{figure*}

\subsection{Statistical Description} 
The proposed dataset consists of around 369K programs written in C/C++/Java/Python programming languages with 809 classes, with each class containing source code submissions written in one or more of C/C++/Java/Python. 
The dataset has been created, keeping 10 as the minimum threshold for the number of source codes for a particular language in each class. Tab.~\ref{tab:statistics} shows the various fields describing the statistics of the dataset, providing a high-level overview of the dataset and the distribution. 

In Fig.~\ref{fig:dataset-distribution}, we show how well the data points are distributed across different classes in the dataset. 
It provides an overview of the number of classes and the number of source codes that each class contains, ultimately describing the distribution of programs across the dataset.
In this graph, we consider the classes that have 200 to 2K data points on the x-axis and the number of classes on the y-axis.
\\

In Fig.~\ref{fig:lang-distribution}, we show the normal distribution of data points with respect to each language in the dataset. 
This graph shows the distribution of datapoints across different languages.
As shown, the maximum number of programs per language in any of the class is kept at a threshold, which is set to 750. 
From the graph, it can be seen that the dataset contains a significant number of well distributed classes with CPP programs.
We also see that the number of submissions in C and Java are lesser when compared to that of CPP.

In Fig.~\ref{fig:tag-distribution}, we show the distribution of the collected code tags for different problems in the dataset. It can be seen that the number of code tags for a class varies from 0 to 8, with the majority of the classes/problems associated with 2 code tags 
in the dataset. 
Classes having two or more tags can be used for NLP-based applications such as code tagging. 
Such classes form a major portion of our dataset.

\section{Applications}
Extending some research questions that we listed in Sec.~\ref{sec:introduction}, we also hope that COFO can be useful in addressing the following broad research areas.

\subsection{Program classification/recognition}
Classifying programs based on the problem statement or detecting if two programs/executables solve the same
purpose is undecidable. 
However, it is a rather important application: classification of programs based on functionality helps in better maintenance of codes, helps in better code suggestion, etc.~\cite{tbcnn-aaai16}.
Solving this problem would need a dataset containing many classes, and each class containing several implementations solving the same problem.
Our proposed dataset would be a natural fit for this task.

\subsection{Code Tagging}
Given a piece of code, predicting a set of relevant tags associated with it, is called as code tagging. 
A tag is a keyword that describes the source code.
As the proposed dataset has a description of the problem in the form of tags, it can be useful in solving this problem.
There are about 35 unique code tags in the proposed dataset, with 2 code tags per problem, on average.

\subsection{NLP-based applications} Other applications can also emerge from the intersection of NLP and source codes. The proposed dataset also contains a problem specification that describes the problem statement. This natural language text, along with the source codes of submissions for that problem, can fall under the category of program comprehension tasks such as code summarization and code search by making use of a modified version of the problem specification, acting as a summary or query for the program. 
Modifying the problem specifications for this task requires some effort, but we believe that the proposed dataset can be used to solve this problem.

\section{Conclusions and Future Work}

We present COFO, a dataset well suited for program classification/recognition and code tagging,  collected from Codeforces, an online platform that hosts competitive programming contests.  COFO has 369K programs (in 809 classes). 
The proposed dataset consists of more classes and programs than the available datasets.
The proposed dataset is orders of magnitude larger than the dataset proposed by Mou et al., containing 50K programs in 104 classes, and is currently a standard dataset used in program classification. 

The dataset contains programs from a wide variety of programming languages, along with the metadata indicating problem and input-output specifications, code tags, and test cases. In the future, we plan to improve the dataset by collecting more data from Codeforces and other similar sources like CodeJam.

\newpage

\bibliography{references}{}
\bibliographystyle{plainurl}

\newpage

\end{document}